\documentclass[twocolumn]{aastex701}

\usepackage{isotope}
\usepackage{amsmath,amssymb}
\usepackage[version=4]{mhchem}
\usepackage{booktabs}

\newcommand{\instr}[1]{\textit{#1}}

\usepackage{color}
\usepackage[dvipsnames]{xcolor}

\graphicspath{{./}{}}

\begin{document}

\title{Production of heavy $\alpha$-elements and $\ce{^44Ti}$ in Cas~A: comparison to abundances from 1D core-collapse supernova models and evidence for Carbon-Oxygen shell mergers}

\author[orcid=0000-0002-4819-310X,sname='Boccioli']{L. Boccioli}
\affiliation{Department of Physics, University of California, Berkeley, CA 94720, USA}
\affiliation{Department of Astronomy, University of California, Berkeley, CA 94720, USA}
\email[show]{lbocciol@berkeley.edu}  

\author[orcid=0000-0003-0390-8770,sname='Roberti']{L. Roberti}
\affiliation{Konkoly Observatory, Research Centre for Astronomy and Earth Sciences, HUN-REN, Konkoly Thege Miklós út 15-17, Budapest, H-1121, Hungary}
\affiliation{CSFK HUN-REN, MTA Centre of Excellence, Konkoly Thege Miklós út 15-17, Budapest, H-1121, Hungary}
\affiliation{Istituto Nazionale di Astrofisica – Osservatorio Astronomico di Roma, Via Frascati 33, Monte Porzio Catone, I-00040, Italy}
\affiliation{Istituto Nazionale di Fisica Nucleare - Laboratori Nazionali del Sud, Via Santa Sofia 62, Catania, I-95123, Italy}
\email[]{}  

\author[0000-0003-2624-0056,sname='Fryer']{C. Fryer}
\affiliation{Center for Theoretical Astrophysics, Los Alamos National Laboratory, Los Alamos, NM 87545, USA}
\affiliation{Center for Nonlinear Studies, Los Alamos National Laboratory, Los Alamos, NM 87545, USA}
\email[]{}  

\author[0000-0001-6189-7665,sname='Safi-Harb']{S. Safi-Harb}
\affiliation{Department of Physics \& Astronomy, University of Manitoba, Winnipeg, MB R3T 2N2, Canada}
\email[]{}  

\author[0000-0003-3970-1843,sname='Jones']{S. Jones}
\affiliation{Theoretical Division, Los Alamos National Laboratory, Los Alamos, NM 87545, USA}
\email[]{}  



\author[orcid=0000-0002-9048-6010,sname='Pignatari']{M. Pignatari}
\affiliation{Konkoly Observatory, Research Centre for Astronomy and Earth Sciences, HUN-REN, Konkoly Thege Miklós út 15-17, Budapest, H-1121, Hungary}
\affiliation{CSFK HUN-REN, MTA Centre of Excellence, Konkoly Thege Miklós út 15-17, Budapest, H-1121, Hungary}
\affiliation{Bavarian Geoinstitute (BGI), University of Bayreuth, Universitätsstraße 30, 95447 Bayreuth, Germany}
\affiliation{NuGrid Collaboration, http://www.nugridstars.org}
\email[]{}

\begin{abstract}
    The merger between the carbon (C) and oxygen (O) shells hours to days before the collapse of a massive star significantly changes its nucleosynthesis, which is reflected in the elemental ratios observed in supernova remnants (SNRs). We present a nucleosynthesis study of $^{44}$Ti production in core-collapse supernovae (CCSNe), highlighting large silicon (Si), sulfur (S), calcium (Ca), and, most importantly, argon (Ar) to neon (Ne) ratios as diagnostics for carbon-oxygen (C--O) shell mergers. We compare yields from eight different sets of CCSNe models to observations of Cassiopeia A (Cas A), and show that C--O shell mergers are consistently the models that best match X-ray and infrared observations. These models produce high Ar/Ne ratios ($\gtrsim 0.1$), due to $^{20}$Ne depletion and production of $^{36}$Ar and $^{38}$Ar, while lower ratios are obtained from non-merger cases. Based on the Ar/Ne diagnostic, we compare the range of expected $^{44}$Ti produced by C--O shell mergers, which is up to $\sim 20 - 30 \%$ of the overall $^{44}$Ti, but expected to be located outside the reverse shock. Based on the sets of models considered, the photon flux expected from the $^{44}$Ti synthesized in the C--O shell merger in Cas A is below the \instr{NuSTAR} and \instr{COSI} detection limits, compatible with current limits locating most of the $^{44}$Ti interior to the reverse shock, but might be detectable from proposed missions like \instr{ASCENT}. Finally, for the SNR of 1987A, a dominant C--O merger origin of the observed $^{44}$Ti is unlikely based on the observed redshift in its $^{44}$Ti line.
\end{abstract}



\section{Introduction}
During the explosion of a core-collapse supernova (CCSN), the shock travels through the outer layers of the star, compressing and heating up material that is eventually swept up and ejected. Some of this material reaches temperatures high enough to synthesize new elements during successive explosive burning episodes \citep{Arnett1969_ExplNuclStar,Woosley1995_EvolExplMassStar2ExplHydr,Limongi2018_PresEvolExplNuclRotaMassStar}. The rest of the material is instead ejected without being reprocessed, and all of the explosive and pre-supernova nucleosynthetic signatures will therefore leave an imprint in the supernova remnant (SNR) hundreds of years after the explosion \citep{Milisavljevic2017_SupeSupeRemnConna,Jerkstrand2026_CoreSupe}.

One of the most studied SNRs is Cassiopeia A (Cas~A), which has recently been suggested to be the result of the explosion of a star that experienced a C--O shell merger \citep{Sato2025_InhoStelMixiFinaHourCassSupe}. The C--O shell merger is a phenomenon observed in several 1D stellar evolution simulations \citep{Ritter2018_NuGrStelDataSetIIStelYiel, Roberti2025_OccuImpaCarbShelMergMassStar} and has also been studied with multi-dimensional simulations of convective shells in massive stars \citep{Andrassy2020_3DHydrSimuIngeConvShel, Rizzuti2023_3DStelEvolHydrSimuCompBurn, Rizzuti2024_ShelMergLateStagMassStarEvol}. Days or even hours before core-collapse, the O-burning shell can ingest material from the overlying Ne and C shells, activating several fusion reactions involving $\ce{^12C}$, $\ce{^16O}$ and, as a consequence, several proton-capture reactions predominantly on S, P, and Ar \citep[i.e. SPAr burning,][]{Roberti2025_SPArBurnProtCaptPoweCarbShel} powering the convective-reactive event.

C--O shell mergers have been investigated mostly because of their nucleosynthetic signatures in the production of p-process nuclei beyond Fe \citep[][]{Ritter2018_NuGrStelDataSetIIStelYiel, Roberti2023_GproNuclCoreSupeNoveAnalGpro} and odd-Z nuclei \citep{Issa2025_3DMacrPhysLighOddzElemProd, Roberti2025_OccuImpaCarbShelMergMassStar, Issa2026_Impa3DMacrNuclPhysP-nuOC}. In light of these studies, the \instr{XRISM} collaboration \citep{Audard2025_ChloPotaEnriCassSupeRemn,Sato2025_InhoStelMixiFinaHourCassSupe} has recently analyzed the emission from Cl, K, Ne, Mg, Si, and Ar in Cas~A, finding many similarities with the nucleosynthetic signatures of C--O shell merger progenitors. 

In this work, we study the production of $\ce{^44Ti}$ in the C--O merger, together with the $\alpha$-elements neon (Ne), silicon (Si), sulfur (S), argon (Ar), and calcium (Ca), and directly compare it to abundance estimates of Cas~A and other remnants. Contrary to previous studies, which focused on selected regions in the remnant, we focus on the entire observable ejecta. Therefore, we are able to derive a global and robust condition to determine the occurrence of a C--O merger based on the Ar/Ne ratio estimated in any SNR. Moreover, we also discuss predictions and place bounds on the amount of $\ce{^44Ti}$ in Cas~A that could have been produced in a C--O shell merger.

This work is organized as follows: in Section~\ref{sec:methods} the stellar models adopted are described; in Section~\ref{sec:merger_nucleosynthesis} we summarize the basic nucleosynthesis in C--O shell mergers; the production of $\alpha$-elements and of $\ce{^44Ti}$ in comparison with Cas~A observations are given in Sections~\ref{sec:alpha_eles_SNR} and~\ref{sec:Ti_Cas-A}, respectively, where we also discuss how velocity distributions in the ejecta of 1987A can be used as indications of C--O shell mergers. Summary and conclusions are in Section~\ref{sec:conclusions}.


\section{Methods}
\label{sec:methods}

To investigate the impact of C--O mergers on the production of heavy $\alpha$-elements and $\ce{^{44}Ti}$, we analyze the explosive nucleosynthesis yields from eight different sets of pre-supernova models available in the literature. The explosion and nucleosynthesis of seven of these sets (for a total of 156 progenitors) were computed following \citet{Boccioli2025_ExplMattHowRealNdriExplChan}. These are 1D+ explosion simulations, carried out using the open-source code \textsc{GR1D} \citep{OConnor2011_BlacHoleFormFailCoreSupe,OConnor2015_OpenRadiHydrCodeCoreSupe} coupled with the neutrino-driven convection model \textit{STIR} \citep{Couch2019_SimuTurbNdriCoreSupeExplOne,Boccioli2021_GeneRelaNdriTurbOnedCoreSupe}. Contrary to other 1D explosion simulations, \textsc{GR1D+} uses a time-dependent mixing-length theory (MLT) prescription calibrated to multi-dimensional simulations. Therefore, explosion energies, mass cuts, and explodability arise consistently once the mixing length has been calibrated. To these, we added 118 CCSN yields from \cite{Sukhbold2016_CoreSupe9120SolaMassBase}, whose explosion simulations follow a more approximate prescription. However, as we will show throughout the paper, most of the material coming from the C--O shell ends up being ejected with very little reprocessing from the shock, showing that this analysis is only weakly dependent on the exact prescription adopted for the explosion, and it is mostly dependent on the pre-supernova evolution. We selected progenitor models spanning a wide range of initial conditions, and more details about the pre-supernova and explosion models can be found in Appendix~\ref{app:preSN_models}. Out of these 274 progenitor models analyzed, 20 show evidence of a C--O shell merger.

Although 1D CCSN models are widely used for various astrophysical applications -- such as input for galactic chemical evolution simulations and direct comparisons with observations (e.g., stellar archaeology and cosmochemistry) -- their limitations in reproducing certain observational features have been previously highlighted \citep[see, e.g.,][for comparisons with SNRs]{Braun2023_ProgExplPropSupeRemnHostCent, Orlando2025_FilaEjecNetwCassReveFingSupe, Orlando2025_TracEjecStruSupe1987InsiDiag, Treyturik2026_ReviSupeEngi3C397W49BSupe}. Nevertheless, extensive grids of 1D CCSN models enable building a broad picture of abundance yields and nuclear reaction rate combinations achievable across progenitor histories \citep[e.g.,][]{Roberti2025_OccuImpaCarbShelMergMassStar} and explosion configurations \citep[e.g.,][]{Andrews2019_NuclYielCoreSupeProsNextGene}, within a realistic range of conditions. 

It should be stressed that 1D models may not accurately predict the integrated yields of individual isotopes and their spatial distributions due to multi-dimensional mixing, during both the pre-supernova convection and the explosion phase. However, they still produce similar thermodynamic conditions to 3D simulations since the nuclear reactions driving the production of the elements are expected to be the same \citep[e.g.,][]{Roberti2025_SPArBurnProtCaptPoweCarbShel}. Therefore, specific isotopic and elemental ratios for abundant species sharing the same nucleosynthesis origin are preserved when the same nuclear rates are adopted. 
In the following sections, we demonstrate how we can leverage the advantages in computational efficiency and parameter exploration of 1D models to provide valuable insights into SNRs despite their limitations relative to 3D simulations. More detailed comparisons with 3D simulations, particularly regarding asymmetries and other multi-D effects, are beyond the scope of this paper and left for future work.

\section{Pre-explosive nucleosynthesis in C--O shell mergers and production of $\ce{^{44}Ti}$}
\label{sec:merger_nucleosynthesis}

C--O shell mergers are characterized by the ingestion of C-, Ne-, and O-rich material into the former O-burning shell from the overlaying burning shells. They occur a few hours to a few days before collapse, and their occurrence generally depends on the CO core mass and the central $^{12}$C abundance at the end of core He burning \citep{Roberti2025_OccuImpaCarbShelMergMassStar}. At such late stages, this ingestion provides the fuel to sustain an even more energetic O-burning, leading to a significant overproduction of light particles, such as protons and alpha particles, which can promptly interact with the surrounding environment. Specifically, radiative proton-capture reactions are activated via SPAr burning, which can dominate the local energy production \citep{Roberti2025_SPArBurnProtCaptPoweCarbShel}. The major nucleosynthesis products include heavy $\alpha$-chain elements (Si, S, Ar, and up to Ca), as well as odd-Z elements \citep[P, Cl, K, Sc, ][]{Ritter2018_ConvReacNuclScClPproIsot, Roberti2025_OccuImpaCarbShelMergMassStar, Issa2025_3DMacrPhysLighOddzElemProd}. The freshly synthesized material pollutes both the former O- and C-shell regions and will be ejected largely unprocessed by the subsequent CCSN explosion.

Another isotope that can be produced in C--O shell mergers is $^{44}$Ti \citep{Chieffi2017_Synt44Ti56NiMassStar, Roberti2025_SPArBurnProtCaptPoweCarbShel, Issa2025_PresOCShelMergCoulProdMore}. With a lifetime of $\sim 60$ years, its radioactive decay can be observed in SNRs that are a few hundred years old \citep{Grefenstette2014_AsymCoreSupeMapsRadi44TiCass}. At the same time, this means that only hydrostatic burning at the very end of stellar evolution or explosive nucleosynthesis will be relevant for its production. However, the $^{44}$Ti produced during hydrostatic Si-burning is typically destroyed by the passage of the shock, and only a very marginal production occurs in other hydrostatic burning in classical stellar evolution models. Several production channels are instead available during explosive nucleosynthesis, most notably incomplete Si-burning, $\alpha$-rich freeze out, and the ``(p-$\gamma$)-leakage'' channel in neutrino-driven winds \citep{Magkotsios2010_TrenTi44Ni56CoreSupe}. Traditionally, spherically symmetric explosions
underestimate the production of $\ce{^{44}Ti}$ by a factor of 3--5 compared to Cas~A and 1987A observations \citep{Woosley2007_NuclRemnMassStarSolaMeta, Sukhbold2016_CoreSupe9120SolaMassBase, Limongi2018_PresEvolExplNuclRotaMassStar, Curtis2018_PUSHCoreSupeExplSpheSymmIII, Boccioli2025_ExplMattHowRealNdriExplChan}. However, as shown by the latest multi-dimensional simulations \citep{Sieverding2023_Prod44TiIronNuclEjec3DNdri, Wang2024_InsiProd44TiNickIsotCoreSupe}, a combination of re-heating at low densities due to multi-dimensional turbulent effects, as well as proton-rich neutrino-driven winds, can boost the production of $\ce{^{44}Ti}$ to values typical of SNRs, without requiring overproductions of $\ce{^56Ni}$ as in earlier works \citep{Young2006_ConsProgCass}.

\section{Alpha-elements in SNR} 
\label{sec:alpha_eles_SNR}

SNRs can be observed across a wide range of wavelengths; for this paper, we focus on infrared and X-ray wavelengths. These two bands provide direct access to emission lines from the key elements mentioned above.

\subsection{Infrared observations}
\label{sec:IR}
Instruments such as \instr{Spitzer} and \instr{JWST} can map the entire remnant of Cas~A \citep{Laming2020_ElemAbunUnshEjecCass,Milisavljevic2024_JWSTSurvSupeRemnCass}. They can therefore be used to estimate the amount of unshocked ejecta, with the most recent estimate being $0.47^{+0.47}_{-0.27}\  M_\odot$ \citep{Laming2020_ElemAbunUnshEjecCass}, compatible with other estimates: $0.18 - 0.3\ M_\odot$ from \citet{Hwang2012_ChanXraySurvEjecCassSupeRemn} and $0.39\ M_\odot$ from \citet{DeLaney2014_DENSMASSUNSHEJECCASSLOWFREQ}. The most recent \instr{JWST} observations focused instead on 4 specific points in the shocked and unshocked ejecta. Given the large contribution of dust emission and the theoretical challenges in modeling it, deriving abundance ratios is far from straightforward. Nonetheless, spectra of the unshocked material exhibit strong Ar and S emission lines. Notice that Ne lines in the unshocked material (specifically, in the so-called ``Green-Monster'') have been detected \citep{Milisavljevic2024_JWSTSurvSupeRemnCass}. However, given their low velocities, they might be due to circumstellar gas, and therefore their central location might simply be a projection effect \citep{Milisavljevic2024_JWSTSurvSupeRemnCass}. This overall would favor a C--O merger scenario (see also Section \ref{sec:X-ray}). 

Given the small field size of the regions selected for the NIRCam observations \citep[$\sim 10-50\ {\rm arcsec}^2$][]{Milisavljevic2024_JWSTSurvSupeRemnCass}, a global analysis directly comparing to 1D simulations could be misleading. Nonetheless, these infrared observations show evidence that a C--O merger scenario could be invoked in the context of Cas~A. To demonstrate this, we now compare nucleosynthesis models more quantitatively with X-ray observations.



\subsection{X-ray observations}
\label{sec:X-ray}
The elemental composition of SNRs can be observed by X-ray telescopes like \instr{Chandra} \citep{Weisskopf2000_ChanXrayObseCXOOver}, \instr{XMM-Newton} \citep{Jansen2001_XMMNObseSpacOper}, and  \instr{XRISM} \citep{XRISMScienceTeam2020_ScieXrayImagSpecMissXRIS}. However, only the material outside of the reverse shock (i.e., in the shocked region) is detectable, because the yet unshocked material is too cold to emit X-ray photons. Once the reverse shock hits this material, it will be heated to X-ray-emitting temperatures. Deriving elemental masses and mass ratios from X-ray observations depends on several assumptions adopted for the plasma modeling. Therefore, different groups find different values for the overall masses of each element, as well as mass ratios, which is why we focus on overall trends and global comparisons instead of trying to match a specific stellar model to a specific observation. 

\begin{figure*}
    \centering
    \includegraphics[width=\linewidth]{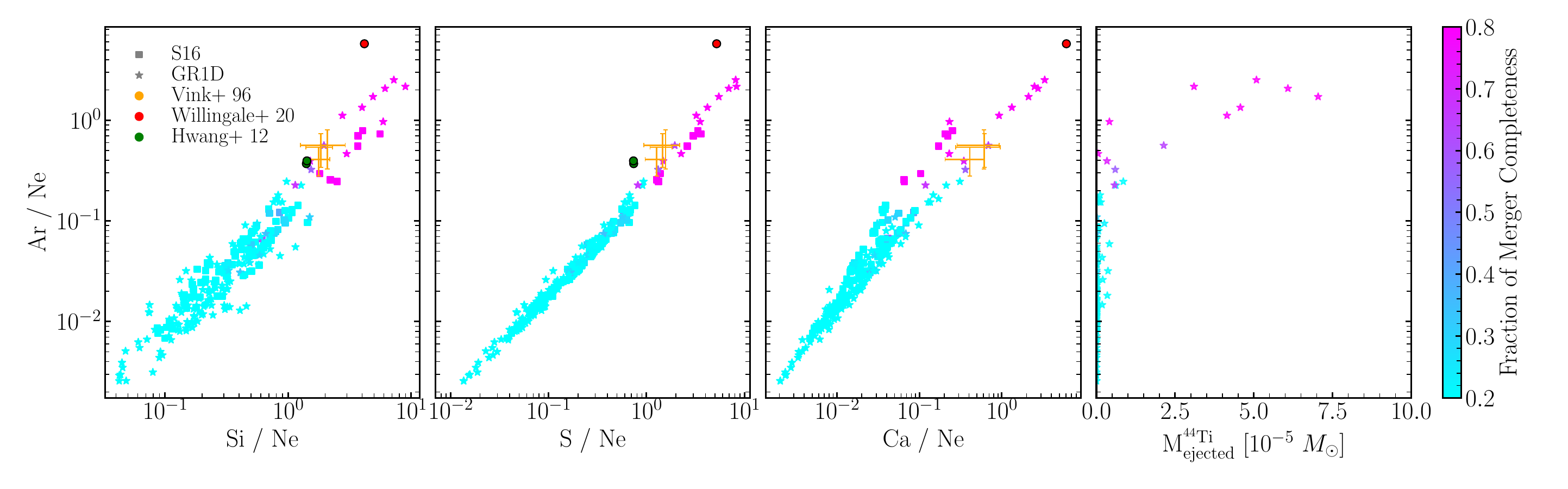}
    \caption{Mass ratios of selected elements after the explosion, when all isotopes have decayed back to stability, for the nucleosynthesis yields from \citet{Boccioli2025_ExplMattHowRealNdriExplChan} (star symbols) and \citet{Sukhbold2016_CoreSupe9120SolaMassBase} (squares). The x-axis of the rightmost panel shows the amount of $\ce{^44Ti}$ synthesized in the C--O merger and then ejected, calculated as outlined in the caption of Figure~\ref{fig:Ti44_ej_frac}. Points are color-coded by the fraction of merger completeness (i.e. hoe well mixed the shell is \citep{Loddo2026_}), where everything below $\sim 0.4-0.5$ (i.e. the cyan points) can be considered to not have experienced a C--O merger. The green dots show mass ratio estimates from \citet{Hwang2012_ChanXraySurvEjecCassSupeRemn} for two values of the filling factor. Orange crosses are from three different regions (N, SE, W) observed by \citet{Vink1996_NewMassEstiPuzzAbunSNRCass}. Red points are from \citet{Willingale2002_XraySpecImagDoppMappCass}, but since the rms uncertainty associated with them is very large, we decided not to show it.}
    \label{fig:ArNe_SiNe_Ti44}
\end{figure*}

In this analysis, we focus on observations and abundance estimates from \citet{Vink1996_NewMassEstiPuzzAbunSNRCass, Willingale2002_XraySpecImagDoppMappCass} and \citet{Hwang2012_ChanXraySurvEjecCassSupeRemn}. We also compared with the abundance ratios of \citet{Sato2025_InhoStelMixiFinaHourCassSupe} who, however, only focused on specific regions, spanning a relatively small portion of the remnant. As a consequence, the mass ratios reported for their investigated regions span the entire range of x and y axes in the two leftmost panels of Figure~\ref{fig:ArNe_SiNe_Ti44}, which the authors attribute to an inhomogeneous mixing during the C--O merger. However, only detailed 3D simulations of the convective-reactive merger could investigate spatial distributions and inhomogeneities in the ejecta, and we therefore do not speculate further in this work. 

As briefly outlined in the previous section \citep[see also][for more details]{Milisavljevic2024_JWSTSurvSupeRemnCass}, the composition of unshocked ejecta probably reflects Si- and O-burning conditions, given the presence of O, Si, S, and Ar \citep{Thielemann1996_CoreSupeTheiEjec, Chieffi1998_Evol25StarMainSequOnseIron, Woosley2002_EvolExplMassStar}. Therefore, the estimates of these elements' total yields based on X-ray observations are most likely a lower limit. 

\subsection{Model comparison to observations}
\label{sec:comparison_alphas}
In progenitors experiencing C--O shell mergers, extended parts of the SNR are expected to be enriched with heavy $\alpha$-elements synthesized by the O burning. In particular, as discussed in Section~\ref{sec:merger_nucleosynthesis}, the C--O shell merger overproduces Si (\isotope[28,29,30]{Si}), S (\isotope[32,34]{S}), Ar (\isotope[36,38]{Ar}), and possibly Ca (mostly \isotope[40,42]{Ca}), while it tends to destroy C-burning products (such as \isotope[20]{Ne}, \isotope[23]{Na}, and \isotope[24]{Mg}). In a typical pre-SN progenitor, Ne and Mg are not reprocessed by the shock, since they are usually synthesized in the C-burning shell, where peak temperatures are quite low. Therefore, potentially, they are both suitable elements to use as a reference. However, Ne is more sensitive to its combined depletion by photodisintegration, via the $^{20}$Ne($\gamma$,$\alpha$)$^{16}$O reaction and the $\alpha$-capture reaction $^{20}$Ne($\alpha$,$\gamma$)$^{24}$Mg, directly converting Ne in Mg. This last channel mitigates the Mg depletion in these conditions. Therefore, Ne is more easily depleted in a C--O shell merger, providing a more sensitive diagnostic compared to Mg. 

Recently, several studies have targeted peculiar elements synthesized during the C--O merger and analyzed their abundance ratios in different regions of Cas~A \citep{Audard2025_ChloPotaEnriCassSupeRemn,Sato2025_InhoStelMixiFinaHourCassSupe}. In particular, they focused on Cl/S versus K/Ar \citep{Audard2025_ChloPotaEnriCassSupeRemn} and Ne/Mg versus Si/Mg \citep{Sato2025_InhoStelMixiFinaHourCassSupe}. However, from the nucleosynthesis point of view, most of these ratios may not provide a robust diagnostic to verify whether the progenitor of the SNR experienced a C--O merger. For example, the odd elements Cl and K are both expected to be strongly overproduced by the C--O merger \citep[e.g.,][]{Ritter2018_ConvReacNuclScClPproIsot}. However, Cl/S and K/Ar ratios also suffer from the S and Ar enhancement, and therefore, they may also be more sensitive to local dynamical properties that may not be accurately captured by 1D models. The Ne/Mg ratio is also not expected to be a robust diagnostic, since both Ne and Mg are destroyed at typical O-burning conditions.

An additional benefit of using Ne is that it is not going to be affected by molecular or dust condensation. Similarly, the noble gas Ar is preserved by further chemical processing in SNR material. This can be quite important since, for example, due to the complex modeling of dust formation and its molecular chemistry, IR observations have identified the same lines with different molecules: \citet{Arendt2014_INTEEJECDUSTCASSUPEREMNa} attributed $21\ \mu{\rm m}$ dust features to $\ce{Mg_{0.7}SiO_{2.7}}$, whereas the same features have been attributed to
$\ce{SiO_{2}}$ grains by \citet{Rho2018_DustTwinCasCoolDust21Mm}. Therefore, mass estimates in the ejecta of Si and Mg become even more uncertain, whereas Ne and Ar overcome this problem entirely.


In Figure \ref{fig:ArNe_SiNe_Ti44}, we show our diagnostic ratios to distinguish progenitors with a C--O shell merger from those without. In the three-element plots shown, the combined high Ar/Ne, Si/Ne, S/Ne, and Ca/Ne ($\gtrsim 0.1$, $\gtrsim 1$, $\gtrsim 1$, and $\gtrsim 0.2$ respectively) allow us to discriminate between models with and without a C--O shell merger. Notice that Ca/Ne is not as good a diagnostic as the other ratios, as shown by the overlap between progenitors with and without C--O mergers with similar Ca/Ne ratios. This is due to the significant explosive component of Ca, which is also the cause of the slightly different slopes found in the \textsc{GR1D} models and in the ones from \citet{Sukhbold2016_CoreSupe9120SolaMassBase}. We also include data points where mass ratios were derived by \cite{Vink1996_NewMassEstiPuzzAbunSNRCass,Willingale2002_XraySpecImagDoppMappCass,Hwang2012_ChanXraySurvEjecCassSupeRemn} for Cas~A. As mentioned above, different abundance estimates from different authors can change the value of these ratios. Nonetheless, the overall trends can be clearly seen across observations, and one can conclude that the only models matching observational estimates for all the three-element plots in Figure \ref{fig:ArNe_SiNe_Ti44} are those that underwent a C--O shell merger.


\section{$^{44}$Ti in Cas~A}
\label{sec:Ti_Cas-A}



Another powerful diagnostic in SNR is radioactive decay of $\ce{^44Ti}$. The main reason is that the plasma and spectral modeling uncertainties mentioned above do not affect $\ce{^44Ti}$, since decay lines are usually a much cleaner signal to detect by hard X-ray and gamma-ray telescopes, such as \instr{NuSTAR} and \instr{COSI}, respectively. This means that, in principle, $\ce{^44Ti}$ can be observed everywhere in the remnant. The latest estimate of the amount of $\ce{^44Ti}$ produced by Cas~A is $(1.54 \pm 0.21) \times 10^{-4} M_\odot$ \citep{Grefenstette2014_AsymCoreSupeMapsRadi44TiCass,Grefenstette2016_DistRadi44TiCass}. Its spatial distribution has been mapped by \cite{Grefenstette2014_AsymCoreSupeMapsRadi44TiCass} using \instr{NuSTAR} data \citep{Harrison2013_NuclSpecTeleArraNuSTHighXRay}, revealing that most of the $\ce{^44Ti}$ is centrally located.

\begin{figure}
    \centering
    \includegraphics[width=\linewidth]{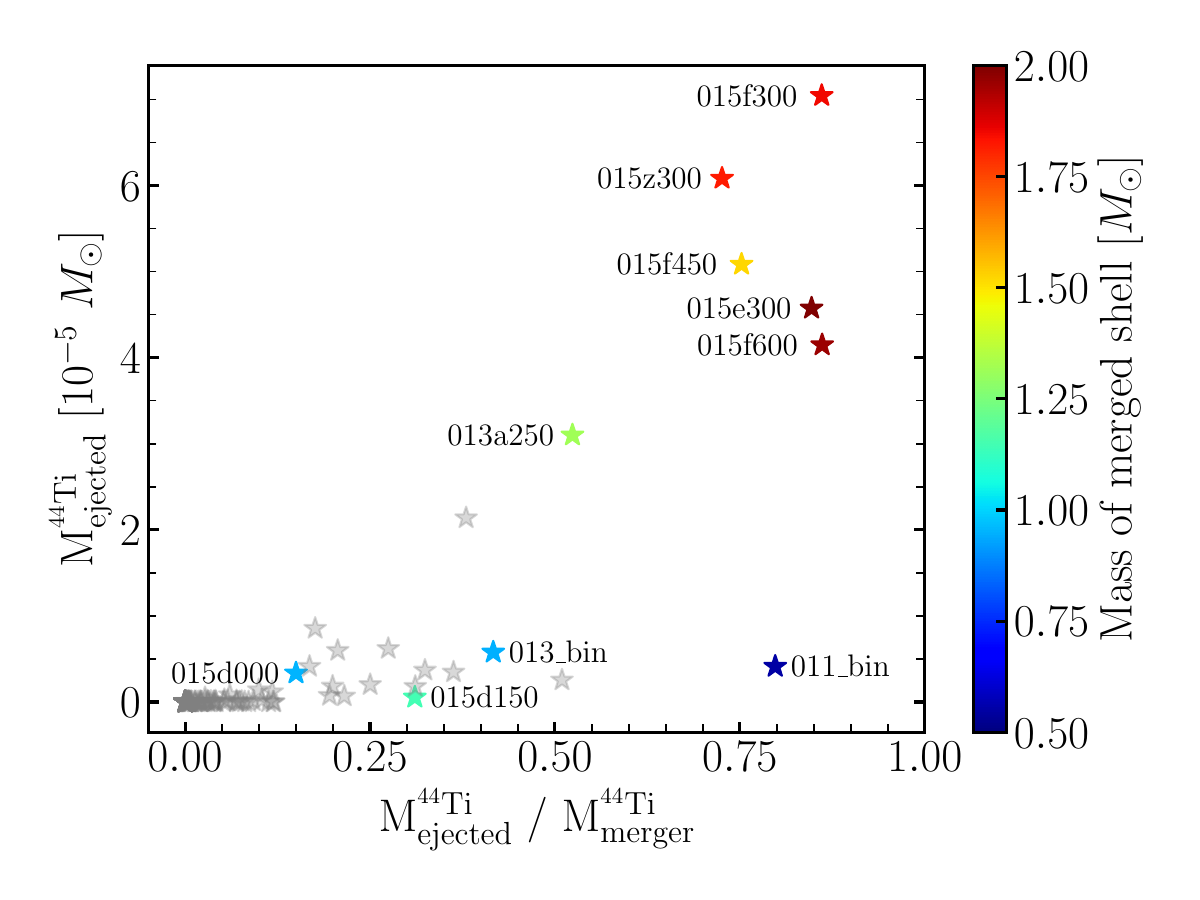}
    \caption{M$^{\rm ^{44}Ti}_{\rm merger}$ represents the $\ce{^44Ti}$ produced in the merger. We define it as the integrated yield above the Si/Si-O interface in the pre-SN progenitor. M$^{\rm ^{44}Ti}_{\rm ejected}$ is then calculated by considering only the shells above the Si/Si-O interface where the pre-SN abundance of $\ce{^44Ti}$ changes by less than 1~\% after the explosion. Grey points are stars without a C--O merger, whereas those with a C--O merger have been color-coded by the mass of the merged shell, and annotated with the model ID convention outlined in Table~\ref{tab:merger_description} and in Appendix~\ref{app:preSN_models}.}
    \label{fig:Ti44_ej_frac}
\end{figure}

This seems to be at odds with the C--O shell merger scenario, according to which some $\ce{^44Ti}$ should be present in O-rich regions located in the shocked regions farther out. One possibility is that the $\ce{^44Ti}$ produced in the C--O merger is destroyed by the passage of the shock. However, as shown in the right panel of Figure~\ref{fig:ArNe_SiNe_Ti44} and in Figure~\ref{fig:Ti44_ej_frac}, a significant fraction is still ejected, and we will discuss what the expected photon flux from this component would be in Section~\ref{sec:photon_flux}. When the sudden merger ignites nuclear burning, the shells expand and decrease their densities, readjusting the envelope to respond to the newly ignited burning episode. Later on, when the shock sweeps the shell, this causes the peak temperatures to decrease relatively quickly. The net result, as shown in Figure~\ref{fig:Ti44_ej_frac}, is that most of the C--O shell and $\ce{^44Ti}$ synthesized in the merger are ejected without being reprocessed by any explosive burning.

The total amount of $\ce{^44Ti}$ produced in the C--O merger that is eventually ejected (i.e. y-axis in Figure~\ref{fig:Ti44_ej_frac}) ranges from $5 \times 10^{-6} - 5 \times 10^{-5}\ M_\odot$, for models with Ar/Ne ratios compatible with Cas~A. This corresponds to $\sim 5-30 \%$ of the total Cas~A yield. Notice that, since 1D CCSN models typically underestimate the explosive $\ce{^44Ti}$ production by a factor of 3-5 as discussed earlier, we do not discuss the explosive component, which is however still expected to dominate the overall abundance of $\ce{^44Ti}$.



\subsection{Predicted line fluxes for $\ce{^44Ti}$ }
\label{sec:photon_flux}

In \citet{Grefenstette2016_DistRadi44TiCass}, \instr{NuSTAR} data is used to infer the amount of $\ce{^44Ti}$ in the remnant. The angular resolution is $18"$ and, since Cas~A spans $\sim 5$~arcmin in the sky, the remnant is mapped in 64 different regions. Only in 10 of these bins (all centrally located), a positive detection of the 68 keV line was reported. Upper limits for the photon flux $F_{68}$ were provided for the other regions \citep[see Figure 7 of][]{Grefenstette2016_DistRadi44TiCass}, ranging from $5-10 \times 10^{-7}\ \gamma\ {\rm cm}^{-2}\ {\rm s}^{-1}$. 

To test whether the $\ce{^44Ti}$ ejected from the C--O shell merged shell could be detected, we calculated the expected photon flux $F_{68}$ by using Equation (1) from \citet{Weng2024_UppeLimi44TiDecaEmisFourNear}:
\begin{equation}
    M(\ce{^44Ti}) = 44 m_{\rm p} \times 4 \pi D^2 \left(\frac{\tau}{\rm ln 2}\right) 2^{t/\tau} \times \frac{F_{68}}{I_\gamma},
    \label{eq:M_to_F}
\end{equation}
where $M(\ce{^44Ti})$ is the total mass of $\ce{^44Ti}$, $m_{\rm p}$ is the proton mass, $D$ is the distance of the SNR, $\tau$ is the half life of $\ce{^44Ti}$, $t$ is the age of the SNR, $I_\gamma$ is the absolute $\gamma$ intensity, and $F_{68}$ is the 68 keV line flux. We adopt standard values for $D = 3.4$~kpc and $t = 355$~yrs, as in \citep{Grefenstette2016_DistRadi44TiCass}, and $\tau = 59.1$~yrs and $I_\gamma = 0.93$ as in \citet{Chen2011_NuclDataShee44}. For the amount of $\ce{^44Ti}$, we use $M(\ce{^44Ti}) = 3-3.5 \times 10^{-5}\ M_\odot$ based on the $\ce{^44Ti}$ ejected in models that reproduce Ar/Ne found in Cas~A (see right panel of Figure~\ref{fig:ArNe_SiNe_Ti44}). From Equation~\eqref{eq:M_to_F}, this leads to $F_{68} \approx 4 \times 10^{-6}\ \gamma\ {\rm cm}^{-2}\ {\rm s}^{-1}$.

However, we also have to take into account how the $\ce{^44Ti}$ produced in the merger is distributed. Since it will be spread out over the ejecta of the former C--O shell, it will be distributed on multiple bins of \instr{NuSTAR}, and therefore the flux estimates above (i.e., $M(\ce{^44Ti})$ in Equation~\eqref{eq:M_to_F}) should be divided by the number of bins that span the outer ejecta. A reasonable estimate, based on the maps reported in \citet{Grefenstette2016_DistRadi44TiCass}, would be $\sim 8-10$ bins, i.e. the bins immediately outside the reverse shock. This would lower the expected flux to an average of $\sim 1-4 \times 10^{-7}\ \gamma\ {\rm cm}^{-2}\ {\rm s}^{-1}$, which is a factor of 2-10 below the detection limit of \instr{NuSTAR} mentioned earlier. 

In the future gamma-ray telescope \instr{COSI}, the detection sensitivity of the total amount of $\ce{^44Ti}$ will be a factor of two better than \instr{NuSTAR}, i.e. $1-2 \times 10^6\ \gamma\ {\rm cm}^{-2}\ {\rm s}^{-1}$ \citep{Yoneda2025_EnhaCompTeleImagMaxiPostEsti}. However, the angular resolution of gamma-ray telescopes is much lower, and the entire Cas~A remnant will be mapped into one pixel. Using again \eqref{eq:M_to_F}, this sensitivity corresponds to $\sim 10^{-5} M_\odot$, and \instr{COSI} will therefore be able to significantly improve the estimate for the total amount of $\ce{^44Ti}$ in Cas~A.

Future proposed missions like \instr{ASCENT} \citep{Kislat2023_ASCEBallHardXrayImagSpecTele} could reach narrow line sensitivities of $\sim 2-4 \times 10^{-7}\ \gamma\ {\rm cm}^{-2}\ {\rm s}^{-1}$, and therefore might be able to put better constraints on how much $\ce{^44Ti}$ was produced in Cas~A from the C--O merger.
 
We can also use Equation~\eqref{eq:M_to_F} to derive a conservative upper limit based on non-detection of $\ce{^44Ti}$ in the external regions: even assuming that all those upper limits were close to the $\ce{^44Ti}$ actually present in the bins, they would sum up to $4 \times 10^{-5}\ M_{\odot}$, which corresponds to about 26\% of the total $\ce{^44Ti}$ measured by~\instr{NuSTAR} in Cas~A. 

In summary, we cannot exclude that, if the progenitor of Cas~A experienced a C--O shell merger before collapse, there could still be up to $\sim$ 25\% of $\ce{^44Ti}$ in the outer ejecta, below \instr{NuSTAR} detection limits, although future confirmed (\instr{COSI}) and proposed (\instr{ASCENT}) missions will be able to improve on these estimates. Based on the approximate estimation provided here using 1D models with similar Ar/Ne ratios to what is observed in Cas~A, the fraction of $^{44}$Ti present in the C--O shell would be up to 20\% of the total $^{44}$Ti, which is compatible with the upper limits derived above.

\subsection{1987A: preliminary considerations about C--O shell merger contributions using $\ce{^44Ti}$}
\label{sec:1987A}

1987A is the only other SNR from a core-collapse explosion where a confirmed detection of $\ce{^44Ti}$ lines was reported \citep{Boggs2015_44TiGrayEmisLineSN19ReveAsym}. There, the total $^{44}$Ti mass is measured to be $1.5^{+0.3}_{-0.3} \times 10^{-4}\,M_\odot$, similar to that of Cas~A. We choose not to discuss any elemental mass ratios for this SNR, since most of the X-ray emission from SN1987 is dominated by the circumstellar medium (CSM) \citep{Ravi2024_LateEvolXRayRemnSN1987Inne,XRISMCollaboration2025_TherKinePropEjecSN19ReveXRIS}, and this makes their comparison with stellar models highly uncertain at the moment. More extensive observations are needed.

The breadth and asymmetry in the $\gamma-$ray lines of $^{56}$Ni decay~\citep{Tueller1990_ObseGammLineProfSN1987}, led to the development of the current convection-enhanced neutrino-driven supernova paradigm with a single large ejecta mode predominantly moving away from our line of sight \citep{Grant1993_AnalNuclGammLineProfSN1987}. 
The $^{44}$Ti decay line features are also redshifted \citep{Boggs2015_44TiGrayEmisLineSN19ReveAsym}, suggesting the bulk of the $^{44}$Ti was produced together with $^{56}$Ni. In the case that a high fraction of $^{44}$Ti was produced in a hypothetical extended C--O shell merger, its ejecta component should have been visible with non-redshifted or even blueshifted features. This would have generated a much broader line with no clear redshift. With current observational errors, it is difficult to derive stringent quantitative constraints. Therefore, we conclude that a strong $^{44}$Ti contribution from a C--O merger in 1987A is extremely unlikely.

\section{Conclusions}
\label{sec:conclusions}

In this work, we have conducted a nucleosynthesis investigation for the production of $^{44}$Ti and the $\alpha$-elements Ne, Si, S, Ar, and Ca in core-collapse supernovae (CCSNe), with a focus on the diagnostic potential of large argon-to-neon (Ar/Ne) ratios to identify pre-explosion carbon–oxygen (C--O) shell mergers. By comparing yields from several distinct sets of 274 CCSN models - spanning standard single-star progenitors and those incorporating C--O shell mergers - we demonstrate that merger scenarios provide the most consistent match to the observational properties of Cassiopeia A (Cas~A). 

The key findings are as follows:
\begin{itemize}
\item C--O shell merger models produce high Ar/Ne mass ratios ($\gtrsim 0.1$) through substantial depletion of Ne via nuclear reactions during the merger, while Ar (in particular $^{36}$Ar and $^{38}$Ar) is enhanced. In contrast, non-merger models yield systematically show lower ratios within a wide range of progenitor masses, metallicities, rotations, and CCSN explosion setups considered. We therefore conclude that Ar/Ne is a powerful, observationally accessible tracer for merger signatures in SNRs. Contrary to other elements and elemental ratios previously used to identify C--O merger signatures in Cas~A, Ne and Ar are noble gases, and therefore not affected by molecular or dust condensations in SNRs, and they are accessible to both X-ray (e.g., \instr{\instr{Chandra}}, \instr{XMM-Newton}, \instr{XRISM}) and IR (e.g., \instr{Spitzer}, \instr{JWST}) spectroscopy, of shocked and unshocked ejecta. Combined with the benchmark Ar/Ne ratio, we show that high Si/Ne, S/Ne, and Ca/Ne ratios can also be used to identify the nucleosynthesis signature of C--O shell mergers.

\item We confirm that nucleosynthesis induced by C--O mergers can efficiently produce $^{44}$Ti. This merger component of $^{44}$Ti is expected to have broader velocity distributions and to be located at relatively large radii. However, for the specific case of Cas~A, \instr{NuSTAR} finds $^{44}$Ti to be very centrally located \citep{Grefenstette2016_DistRadi44TiCass}. We therefore derive an observational upper limit for the $^{44}$Ti based on the non-detection in the external Cas~A bins, equivalent to about 20\% of the observed $^{44}$Ti. This number is compatible with 1D model predictions from C--O mergers for models with similar Cas~A Ar/Ne ratios. Instead, based on the very similar redshift in gamma-ray lines and, therefore, velocity distributions of $\ce{^56Ni}$ and $\ce{^44Ti}$, we concluded that a strong $\ce{^44Ti}$ component from a C--O merger in 1987A is extremely unlikely.

\item Similar considerations could be derived for other SNRs, where measurements of Ne and/or Ar are more uncertain or not available \citep[e.g., G15.9+0.2, Kes 79, Puppis A, G349.7+0.2, G350.1–0.3][]{Braun2023_ProgExplPropSupeRemnHostCent}, oftentimes due to their high column density. We therefore reiterate the importance of estimating abundances of both Ne (or, if not available, Mg) and Ar, as these are elements that can robustly probe the occurrence of C--O shell mergers.

\end{itemize}


Although 1D models have inherent limitations in capturing multi-dimensional asymmetries and mixing coming from the intrinsic 3D nature of C--O shell mergers, we expect them to preserve major qualitative signatures such as correlated elemental/isotopic ratios mostly governed by nuclear reaction rates. This makes them valuable tools for diagnostic applications like the Ar/Ne ratio and other ratios explored here (Si/Ne, S/Ne, Ca/Ne). In particular, for the considered mass ratios, we do not see any specific point of failure for 1D CCSN models with the current observational limitations, particularly when considering the entirety of the ejecta.

The large number of 1D models adopted spans a wide range of ZAMS masses, metallicities, and rotations, including some binary progenitors. Therefore, they provide an ideal statistical background to explore the full range of nuclear-reaction activations and ensuing nucleosynthesis signatures to compare with observations, and identify relevant diagnostics for the macroscopic properties of C--O shell mergers. The overall agreement between 1D stellar models, despite their inherent inability to model asymmetries, and observations of the entire ejecta of Cas~A is remarkable. Moreover, their successes and limitations can be used to constrain the main properties of the next generations of 3D C--O shell merger models, with the goal to reproduce quantitative elemental yields and asymmetries in the C--O shell merger of the Cas~A progenitor, or of other SNRs that will be studied in the future.



\begin{acknowledgments}
The authors would like to thank Kasun Wimalasena and Fabian Kislat for useful discussions on ASCENT. NuGrid was supported by the European Union’s Horizon 2020 research and innovation programme (ChETEC-INFRA -- Project no. 101008324), the Lend\"ulet Program LP2023-10 of the Hungarian Academy of Sciences, the Hungarian NKFIH via K-project 138031 and NKKP Advanced grant 153697, and the IReNA network by NSF AccelNet (Grant No. OISE-1927130). LB is supported by the U.S. Department of Energy under Grant No. DE-SC0004658 and SciDAC grant, DE-SC0024388. LB would like to thank the N3AS center for its hospitality and support, and Dan Kasen and Tianshu Wang for fruitful discussions. MP thanks the support from the ERC Synergy Grant Programme (Geoastronomy, grant agreement number 101166936, Germany).  MP thanks access to {\sc viper}, the University of Hull HPC Facility. This research has used the Astrohub online virtual research environment (https://astrohub.uvic.ca), developed and operated by the Computational Stellar Astrophysics group (http://csa.phys.uvic.ca) at the University of Victoria and hosted on the Computed Canada Arbutus Cloud at the University of Victoria. LR acknowledges the support from the PRIN URKA Grant Number \verb|prin_2022rjlwhn|. The work by CLF was supported by the US Department of Energy through the Los Alamos National Laboratory. Los Alamos National Laboratory is operated by Triad National Security, LLC, for the National Nuclear Security Administration of U.S.\ Department of Energy (Contract No.\ 89233218CNA000001).
SSH acknowledges support from the Natural Sciences and Engineering Research Council of Canada (NSERC) through the Canada Research Chairs and the Discovery Grants programs. This work benefited from interactions and workshops co-organized by The Center for Nuclear astrophysics Across Messengers (CeNAM) which is supported by the U.S. Department of Energy, Office of Science, Office of Nuclear Physics, under Award Number DE-SC0023128.
\end{acknowledgments}

\appendix

\section{Pre-supernova and explosion models}
\label{app:preSN_models}

\begin{table}[h!]
\caption{Explosion properties of progenitors that underwent a C--O merger. Notice that none of the progenitors from \citet{Woosley2007_NuclRemnMassStarSolaMeta} and \citet{Limongi2024_EvolFinaFateSolaMetaStarMass} underwent a C--O merger. The first column follows the nomenclature described in Appendix~\ref{app:preSN_models}, the second column shows the reference for each model, where LC18, F23, R24, and F25 refer to \citet{Limongi2018_PresEvolExplNuclRotaMassStar}, \citet{Farmer2023_NuclBinaStar}, \citet{Roberti2024_ZeroExtrLowMetaRotaMassStar}, and \citet{Falla2025_RoleRotaYielTwoGRayEmit26Al}, respectively. Subsequent columns are the final explosion energy $E_{\rm exp}$, the gravitational mass of the cold neutron star $M^{\rm grav}_{\rm NS}$, the baryonic mass of the neutron star (i.e. the mass cut) $M^{\rm bary}_{\rm NS}$, the ejecta mass $M_{\rm ej}$, the pre-SN mass of the model $M_{\rm pre-SN}$, the Ar/Ne ratio, and the mass of $\ce{^44Ti}$ ejected in the C--O shell M$^{\rm ^{44}Ti}_{\rm ejected}$.}
\label{tab:merger_description}
\begin{tabular}{l|cccccccc}
\toprule
Model ID & Ref & $E_{\rm exp}$ & $M^{\rm grav}_{\rm NS}$ & $M^{\rm bary}_{\rm NS}$ & $M_{\rm ej}$ & $M_{\rm pre-SN}$ & Ar/Ne & M$^{\rm ^{44}Ti}_{\rm ejected}$ \\
 & & [$10^{51}$ erg] & [$M_\odot$] & [$M_\odot$] & [$M_\odot$] & [$M_\odot$] &  & [$10^{-5}\ M_\odot$] \\
\midrule
013a300 & LC18 & 0.799 & 1.694 & 1.822 & 3.477 & 5.302 & 0.323 & 0.597 \\
013b150 & LC18 & 0.645 & 1.688 & 1.812 & 9.457 & 11.271 & 0.561 & 2.138 \\
015d000 & LC18 & 0.802 & 1.573 & 1.687 & 13.114 & 14.802 & 0.394 & 0.335 \\
015d150 & LC18 & 1.189 & 1.779 & 1.913 & 11.194 & 13.109 & 0.465 & 0.054 \\
011\_bin & F23 & 0.235 & 1.265 & 1.366 & 1.804 & 3.171 & 0.967 & 0.412 \\
013\_bin & F23 & 0.484 & 1.380 & 1.499 & 2.484 & 3.985 & 0.226 & 0.577 \\
015f600 & R24 & 0.531 & 1.732 & 1.872 & 11.135 & 13.009 & 1.114 & 4.148 \\
015f300 & R24 & 0.502 & 1.732 & 1.871 & 10.758 & 12.630 & 1.712 & 7.044 \\
015z300 & R24 & 0.677 & 1.775 & 1.922 & 4.745 & 6.669 & 2.068 & 6.083 \\
015f450 & R24 & 0.607 & 1.711 & 1.847 & 12.987 & 14.835 & 2.515 & 5.084 \\
015e300 & R24 & 0.536 & 1.799 & 1.948 & 4.504 & 6.454 & 1.339 & 4.574 \\
013a250 & F25 & 0.816 & 1.596 & 1.734 & 3.464 & 5.200 & 2.161 & 3.100 \\
\bottomrule
\end{tabular}
\end{table}

For this analysis, we considered 121 CCSN models by \cite{Boccioli2025_ExplMattHowRealNdriExplChan}, taken from three sets of stellar evolution models. We consider 25 single-star, solar metallicity progenitors from KEPLER \citep{Woosley2007_NuclRemnMassStarSolaMeta}; 57 single-star models at seven different metallicities and 3 different initial rotational velocities from FRANEC \citep{Limongi2018_PresEvolExplNuclRotaMassStar}; 39 solar metallicity progenitors from MESA \citep{Farmer2023_NuclBinaStar}, 16 of which are binary stars and the rest are single stars. To these we added, following the same explosion \citep[simulated with GR1D][]{OConnor2010_NewOpenCodeSpheStelCollNeut, OConnor2015_OpenRadiHydrCodeCoreSupe, Boccioli2021_GeneRelaNdriTurbOnedCoreSupe} and post-process nucleosynthesis setup \citep[with \textsc{SkyNet}][]{Lippuner2017_SkyNModuNuclReacNetwLibr}: 13 rotating, low metallicity progenitors from FRANEC \citep{Roberti2024_ZeroExtrLowMetaRotaMassStar}; 6 low-mass progenitors at solar metallicity from FRANEC \citep{Limongi2024_EvolFinaFateSolaMetaStarMass}; 16 rotating, solar metallicity progenitors from FRANEC \citep{Falla2025_RoleRotaYielTwoGRayEmit26Al}. Finally, we also analyzed the nucleosynthesis yields from 118 exploding models at solar metallicity from \citet{Sukhbold2016_CoreSupe9120SolaMassBase}, where the explosion has been simulated with a more approximate neutrino heating scheme compared to the present work.

The metallicities considered in FRANEC models \citep{Limongi2018_PresEvolExplNuclRotaMassStar, Roberti2024_ZeroExtrLowMetaRotaMassStar, Limongi2024_EvolFinaFateSolaMetaStarMass, Falla2025_RoleRotaYielTwoGRayEmit26Al} range from $z = z_\odot$ to $z = 10^{-6} z_\odot$, as well as some zero metallicity progenitors from \citet{Roberti2024_ZeroExtrLowMetaRotaMassStar}. In this paper, we follow the original FRANEC nomenclature ``xxxmyyy'', where ``xxx'' indicate the zero-age main sequence (ZAMS) mass in units of $M_\odot$, ``yyy'' indicate the initial rotational velocity in units of km/s, and ``m" is a letter referring to the metallicity of the model, with ``a'' being $z = z_\odot$, ``b'' being $z = 10^{-1} z_\odot$, all the way to ``g'' being $z = 10^{-6} z_\odot$, with each subsequent letter scaled down by an order of magnitude each. The letter ``z" refers to zero metallicity progenitors. For the progenitors of \citet{Farmer2023_NuclBinaStar} we instead identify each star with ``xxx\_sin'' or ``xxx\_bin'' depending on wether they are a single or binary star, with ``xxx'' being the ZAMS mass.

From \citet{Sukhbold2016_CoreSupe9120SolaMassBase}, the ZAMS masses of the progenitors that underwent a C--O merger are: 19.7 $M_\odot$, 19.8 $M_\odot$, 20.2 $M_\odot$, 20.4 $M_\odot$, 20.5 $M_\odot$, 25.7 $M_\odot$, and 25.8 $M_\odot$. The other progenitors that underwent a C--O merger and whose explosion we simulated with GR1D \citep{OConnor2010_NewOpenCodeSpheStelCollNeut,OConnor2015_OpenRadiHydrCodeCoreSupe} are listed in Table~\ref{tab:merger_description}, alongside some of the explosion properties.

\bibliography{MyLibrary}{}
\bibliographystyle{aasjournalv7}



\end{document}